\documentclass[twocolumn,preprintnumbers,amsmath,amssymb]{revtex4}
\usepackage{mathrsfs}
\usepackage{amssymb}
\usepackage{graphicx}
\usepackage{dcolumn}
\usepackage{bm}
\usepackage{textcomp}
\usepackage{amsmath}
\usepackage[titletoc]{appendix}

\newcommand\ket[1]{\ensuremath{|#1\rangle}}

\newcommand\oprod[2]{\ensuremath{|#1\rangle\langle#2|}}

\begin{document}
\title{Measurement-device-independent quantum key distribution with source state errors in photon number space}
\author{ Cong Jiang$ ^{1,2}$, Zong-Wen Yu$ ^{1,3}$,
and Xiang-Bin Wang$ ^{1,2,4\footnote{Email
Address: xbwang@mail.tsinghua.edu.cn}\footnote{Also a member of Center for Atomic and Molecular Nanosciences at Tsinghua University}}$}

\affiliation{ \centerline{$^{1}$State Key Laboratory of Low
Dimensional Quantum Physics, Department of Physics,} \centerline{Tsinghua University, Beijing 100084,
People*s Republic of China}
\centerline{$^{2}$ Synergetic Innovation Center of Quantum Information and Quantum Physics, University of Science and Technology of China}\centerline{  Hefei, Anhui 230026, China
 }
\centerline{$^{3}$Data Communication Science and Technology Research Institute, Beijing 100191, China}\centerline{$^{4}$ Jinan Institute of Quantum technology, SAICT, Jinan 250101,
People*s Republic of China}}
\begin{abstract}
The existing decoy-state MDI-QKD theory assumes the perfect control of the source states which is a an impossible task for any real setup. In this paper, we study the decoy-state MDI-QKD method with source errors without any presumed conditions and we get the final security key rate only with the range of a few parameters in the source state.
\end{abstract}


\maketitle

\section{Introduction}
Quantum key distribution (QKD) is one of the most successful applications of quantum information processing. QKD can provide unconditional security based on the laws of quantum physics~\cite{BB84,GRTZ02}. Almost all of the existing setups of QKD use an imperfect single-photon source which, in principle, suffers from the photon-number-splitting (PNS) attack~\cite{PNS,PNS1}. Fortunately, the decoy-state-method~\cite{ILM,H03,wang05,LMC05,wang06,AYKI,peng,wangyang,rep,njp} can help to make a setup with an imperfect single-photon source be as secure as that with a perfect single-photon source~\cite{PNS,PNS1}. Aside from the source imperfection, the limited detection efficiency is another threat to the security~\cite{lyderson}. The device independent QKD (DI-QKD)~\cite{ind1} and the measurement-device-independent QKD (MDI-QKD)~\cite{curty1,ind3} have been proposed to overcome the problem.

The key idea of MDI-QKD is that both Alice and Bob send out quantum signals to the untrust third party (UTP) but neither of them perform any measurement. The UTP would perform a Bell state measurement to each received pule pair and announce whether it's a successful event as well his measurement outcome in the public channel. Those bits corresponding to successful events will be post selected and further processed for the final key. By using the decoy-state method, Alice and Bob can use imperfect single-photon sources~\cite{curty1,wang10} securely in the MDI-QKD. Hence, the decoy-state MDI-QKD can remove all detector side-channel attacks and PNS attacks with imperfect single-photon sources. The combination of the decoy-state method and MDI-QKD has been studied both experimentally~\cite{tittel,chan1,liu1,ferreira,tang1,tang2} and theoretically~\cite{wang10,ind2,a1,a2,a3,a4,a5,a6,a7,a8,a9,a10,a11}.

The existing decoy-state MDI-QKD theory assumes the perfect control of the source states. This is an impossible task for any real setup in practice. As shown in Ref.~\cite{wangyang,njp}, in BB84 decoy-state QKD protocol, the intensity fluctuation, or more generally, the source errors in Fock space may break the equality
\begin{equation}
s_k=s_{k^{'}},
\end{equation}
where $s_k$ is the yield of $k$-photon decoy pulse and $s_{k^{'}}$ is the yield of $k$-photon signal pulse. Similarly to the traditional decoy-state QKD protocol, the source used in the decoy-state MDI-QKD protocol can not be perfectly stable. One important problem is the effect of source errors. In this paper, we would study the decoy-state MDI-QKD method with source errors without any presumed conditions (In most of the case, we have no idea about the details of source error, we can't make any assumptions to the source error model). Only with the range of a few parameters in the source state, we could get the final security key rate in the cost of little decrease. In our method, we have assumed the worst case that Eve knows exactly the error of each pulse. Our result immediately applies to all existing experimental results. Before going further, we emphasize that our results here are unconditionally correct because we have not assumed any unproven conditions. Although there is another approach reported for the issue of intensity error by using the model of attenuation to pulses from an untrusted source, however, there exists counter examples to the elementary equation in that approach, as was shown in the appendix of Ref.~\cite{njp}.     

This paper is arranged as follows. After the Introduction above, we present our method with the virtual protocol and some definitions in Sec.~\ref{virtual} and Sec.~\ref{definitions}. We then show the details in Sec.~\ref{countingrate} and Sec.~\ref{errorrate} about how we formulate the final security key rate with only the bound values of a few parameters in the states involved. And in Sec.~\ref{simulation}, we will show some numerical simulation results. The article is ended with a concluding remark.

\section{Our method}\label{Sec:our}
In the five-intensity decoy-state MDI-QKD protocol, we assume that Alice (Bob) has two sources $v_A$, $x_A$ ($v_B$, $x_B$) in $X$ basis and three sources $w_A$, $y_A$, $z_A$ ($w_B$, $y_B$, $z_B$) in $Z$ basis, and they send pulse pairs (one pulse from Alice, one pulse from Bob) to UTP one by one. Each pulse sent out by Alice (Bob) is randomly chosen from one of the five sources $l_A$ ($r_B$) with constant probability $p_{l_A}$ ($p_{r_B}$) respectively for $l,r=v,x,w,y,z$. Sources $v_A, w_A$ ($v_B, w_B$) are the unstable vacuum sources in $X$ and $Z$ basis respectively. Generally, these unstable vacuum sources are not exact zeros photon-number state. $x_A, y_A$ ($x_B, y_B$) are the decoy sources which are used to estimate the lower bounds of yield and the upper bound of phase-flip error of single-photon pulse pairs. $z_A$ ($z_B$) is the signal source which is used to extract the final key.

We shall use notation $lr$ to indicate the two-pulse source when Alice use source $l_A$ and Bob use source $r_B$ to generate a pulse pair. For simplicity, we omit the subscripts of any $l$ and $r$ for a two-pulse source, eg., source $yz$ is the source that Alice uses source $y_A$ and Bob uses source $z_B$. We also denote the number of counts caused by the two-pulse source $lr$ as $N_{lr}$. $N_{lr}$ are observed values and will be regarded as known values.

\subsection{Virtual protocol}\label{virtual}
For clarity, we first consider a virtual protocol. Suppose Alice and Bob send $N_t$ pulse pairs to UTP in the whole protocol. In photon-number space, the states of the $i$th pulse pair from source $l_A$ ($r_B$) is
\begin{equation}\label{states}
  \rho_{l_A}^{i}=\sum_{k}a_k^{l,i}\oprod{k}{k} \quad(\rho_{r_B}^{i}=\sum_{k}b_k^{r,i}\oprod{k}{k}),
\end{equation}
for $l,r=v,x,w,y,z$. At any time $i (i\in[1,N_t])$, Alice (Bob) choose only one source from $l_A$ ($r_B$) with constant probability $p_{l_A}$ ($p_{r_B}$) respectively. The unselected pulses will be discarded. After UTP has completed all measurements to the incident pulse pairs, Alice (Bob) checks the record about which pulse is selected at each time, i.e., which time has used which source. Obviously, Alice (Bob) can decide which source is to be used at each time in the very beginning. This is just then the real protocol of the decoy-state method. Videlicet, the formulas we get under such a virtual protocol will hold for the real protocol.

\subsection{Some definitions}\label{definitions}
\noindent \textbf{Definition 1.} In the protocol, Alice and Bob send $N_t$ pulse pairs to UTP, one by one. If UTP announces that it's a successful event, then we say that the $i$th pulse pair has caused a count.

Given the source state in Eq.~\eqref{states}, any $i$th pulse sent out by Alice (Bob) must be in a photon number state. We shall make use of this fact that any individual pulse is in one Fock state.

\noindent \textbf{Definition 2.} Set $C$ and $c_{jk}$: Set $C$ contains any pulse pair that has caused a count; set $c_{jk}$ contains any $\ket{jk}$-photon pulse pair (Alice's $j-$photon pulse and Bob's $k-$photon pulse) that has caused a count. Mathematically speaking, the sufficient and necessary condition for $i\in C$ is that the $i$th pulse pair has caused a count. The sufficient and necessary condition for $i\in c_{jk}$ is that the $i$th pulse pair is a $\ket{jk}-$photon pulse pair and it has caused a count. For instance, if the photon number sates of the first 10 pulse pairs sent to UTP are $\ket{ 00 },\ket{ 00 },\ket{ 01 },\ket{ 02 },\ket{ 11 },\ket{ 12 },\ket{ 01 },\ket{ 10 },\ket{ 00 },\ket{ 11 }$, and the pulse pairs of $i=2,3,4,7,8,9$ each has caused a count, then we have
$C=\{i\vert i=2,3,4,7,8,9,\dots\}, c_{00}=\{i\vert i= 2,9,\dots\}, c_{01}=\{i\vert i=3,7,\dots\}, c_{10}=\{i\vert i=8,\dots\}$. Clearly, $C=\cup_{j,k\geq 0} c_{jk}$.

\noindent \textbf{Definition 3.} We use superscripts $U,L$ for the upper bound and lower bound of a certain parameter. In particular, given any $k\ge 0$ in Eq.~\eqref{states}, we denote $a_k^{l,U},a_k^{l,L}$ ($b_k^{r,U},b_k^{r,L}$) for the maximum value and minimum value of $\{a_k^{l,i}\vert i\in C\}$ ($\{b_k^{r,i}\vert i\in C\}$), and $l,r=v,x,w,y,z$. We assume these bound values are known in the protocol.

\subsection{The lower bound of counts of single-photon pulse pairs}\label{countingrate}
Here in this subsection, we only need to consider the pulse pairs in $Z$ basis. According to our definitions, if the $i$th pulse pair is an element of $c_{jk}$, the probability that it is from source $lr$ is
\begin{equation}\label{probability}
  P_{i\vert jk}^{lr}=p_l p_r a_j^{l,i}b_k^{r,i}d_{jk}^i, \quad (l,r=w,y,z),
\end{equation}
where
\begin{equation}
  d_{jk}^{i}=\frac{1}{\sum_{l,r=w,y,z} {p_{l}p_{r} a_{j}^{l,i}b_{k}^{r,i}}}.
\end{equation}
We want to formulate the numbers of $\ket{jk}-$photon pulse pair counts caused by each two-pulse source. Given the definition of the set $c_{jk}$, this is equivalent to asking how many pulse pairs in set $c_{jk}$ come from each two-pulse source. The probability that the $i$th pulse pair ($i\in c_{jk}$) comes from source $l r$ is $P_{i\vert jk}^{lr}$, and equivalently,
\begin{equation}
  n_{jk}^{lr}=\sum_{i\in c_{jk}}P_{i\vert jk}^{lr},
\end{equation}
where $n_{jk}^{lr}$ is the numbers of $\ket{jk}-$photon pulse pair counts caused by source $l r$. Since every pulse pair in $c_{jk}$ has caused a count, therefore we can formulate the total pulse pair counts caused by source $lr$ by
\begin{equation} \label{N_xx}
  N_{lr}=\sum_{j,k\geq 0}n_{jk}^{lr}=\sum_{j,k\geq 0}\sum_{i\in c_{jk}}p_l p_r a_{j}^{l,i}b_{k}^{r,i}d_{jk}^i.
\end{equation}
We also need to introduce the following notation
\begin{align}\label{N_xx1}
  \widetilde{N}_{lr}=\sum_{j,k \geq 1}\sum_{i\in c_{jk}}p_l p_r a_{j}^{l,i} b_{k}^{r,i} d_{jk}^i.
\end{align}
If we define
\begin{equation}
  D_{jk}=\sum_{i\in c_{jk}}d_{jk}^{i},
\end{equation}
our goal as stated in the very beginning of Sec.\ref{Sec:our} is simple to find out the lower bound of $D_{11}$. For, with this and Def.~3, the lower bound of the number of counts caused by those single-photon pulse pairs from the source $l r$, i.e., $n_{11}^{lr}$ is
\begin{equation}
  n_{11}^{lr,L}=p_l p_r a_1^{l,L} b_1^{r,L} D_{11} \leq n_{11}^{lr}.
\end{equation}

In what follows, we shall first find the formula of $D_{11}$ in terms of $\widetilde{N}_{yy},\widetilde{N}_{zz}$ based on Eq.~\eqref{N_xx1}.
\begin{eqnarray}
  \label{N_xxw1}\widetilde{N}_{yy}&=&p_y^2 a_1^{y,U} b_1^{y,U} D_{11}+p_y^2\Lambda-\xi_1,\\
  \label{N_yyw1}\widetilde{N}_{zz}&=&p_z^2 a_1^{z,L} b_1^{z,L} D_{11}+p_z^2\Lambda^{\prime}+\xi_2,
\end{eqnarray}
where
\begin{eqnarray}
  \Lambda &=& \sum_{J}a_j^{y,U} b_k^{y,U} D_{jk},\quad \Lambda^{\prime}=\sum_{J}a_j^{z,L}b_k^{z,L}D_{jk},\\
  \xi_1 &=& p_{y}^2\sum_{j,k\ge 1}\sum_{i\in c_{jk}}(a_{j}^{y,U} b_{k}^{y,U}-a_{j}^{y,i} b_{k}^{y,i})d_{jk}^{i}\ge 0, \\
  \xi_2 &=& p_{z}^2\sum_{j,k\ge 1}\sum_{i\in c_{jk}}(a_{j}^{z,i} b_{k}^{z,i}-a_{j}^{z,L} b_{k}^{z,L})d_{jk}^{i}\ge 0,
\end{eqnarray}
with
\begin{equation}
  J=\{j,k \vert j\ge 1, k\ge 1, jk\ge 2\}.
\end{equation}

Without losing the generality, we assume $K_a=\frac{a_{1}^{z,L}b_{2}^{z,L}}{a_{1}^{y,U}b_{2}^{y,U}} \leq \frac{a_{2}^{z,L}b_{1}^{z,L}}{a_{2}^{y,U}b_{1}^{y,U}}=K_b$ (in the case of $K_a>K_b$, the results can be obtained similarly). According to the definitions of $\Lambda$ and $\Lambda^{\prime}$, we have
\begin{equation}
  \Lambda^{\prime}=K_a \Lambda+\frac{\xi_3}{p_z^2},
\end{equation}
where
\begin{equation}
\frac{\xi_3}{p_z^2}=\sum_{J}(a_j^{z,L}b_k^{z,L}-K_a a_j^{y,U}b_k^{y,U})D_{jk}.
\end{equation}

Further, we assume the important conditions
\begin{equation}\label{decoycondition1}
  \frac{a_k^{z,L}}{a_k^{y,U}}\ge\frac{a_2^{z,L}}{a_2^{y,U}}\ge\frac{a_1^{z,L}}{a_1^{y,U}},\quad
  \frac{b_k^{z,L}}{b_k^{y,U}}\ge\frac{b_2^{z,L}}{b_2^{y,U}}\ge\frac{b_1^{z,L}}{b_1^{y,U}},
\end{equation}
for all $k\geq 2$. The imperfect sources with small error used in practice such as the coherent state source, the heralded source out of the parametric down-conversion, could satisfy the above restriction.

With the assumption $K_a\leq K_b$, one may easily prove that $\xi_3\ge 0$. Thus Eq.~\eqref{N_yyw1} can be rewritten into
\begin{equation}\label{N_yyw11}
  \widetilde{N}_{zz}=p_z^2a_1^{z,L}b_1^{z,L}D_{11}+p_z^2 K_a \Lambda+\xi_2+\xi_3.
\end{equation}
Combining Eqs.~(\ref{N_xxw1},\ref{N_yyw11}), we get the lower bound of $D_{11}$
\begin{equation}\label{conclution1}
  D_{11}\geq D_{11}^L= \frac{\frac{b_2^{z,L}}{p_y^2}\widetilde{N}_{yy}^L-\frac{b_2^{z,L}}{p_z^2 K_a}\widetilde{N}_{zz}^U}{a_1^{y,U}(b_1^{y,U}b_2^{z,L}-b_2^{y,U}b_1^{z,L})},
\end{equation}
where $\widetilde{N}_{yy}^{L}$ and $\widetilde{N}_{zz}^{U}$ are the lower and upper bounds of $\widetilde{N}_{yy}$ and $\widetilde{N}_{zz}$ respectively that will be evaluated in the coming. In obtaining Eq.~\eqref{conclution1}, we have used the facts that $\xi_1,\xi_2$ and $\xi_3$ are all nonnegative values.

In what follows, we shall formulate the lower bound of $\widetilde{N}_{yy}$ and the upper bound of $\widetilde{N}_{zz}$. These bounds can be easily obtained if we assume that Alice and Bob can prepare the vacuum source. However, in practice, the different intensities are usually generated with an intensity modulator, which has a finite extinction ratio. So it is usually difficult to create a perfect vacuum state in decoy-state QKD experiments. In the following of this paper, we will show that these bounds can also be formulated without using the perfect vacuum source.

Similarly to the conditions in Eq.~\eqref{decoycondition1}, we also assume
\begin{equation}\label{decoycondition2}
  \frac{a_k^{l,i}}{a_k^{w,i}}\ge\frac{a_1^{l,i}}{a_1^{w,i}}, \quad \frac{b_k^{r,i}}{b_k^{w,i}}\ge\frac{b_1^{r,i}}{b_1^{w,i}},\quad (l,r=y,z)
\end{equation}
for all $i\in C$. The lower bound of $\widetilde{N}_{yy}$ and the upper bound of $\widetilde{N}_{zz}$ can be expressed by
\begin{equation}\label{N_xxL}
  \widetilde{N}_{yy}^L=\frac{N_{yy}-\widetilde{n}_{yy}^{w,U}}{1+\sigma_C^y}, \quad \widetilde{N}_{zz}^U=\frac{N_{zz}-\widetilde{n}_{zz}^{w,L}}{1-\sigma_A^{z}-\sigma_B^{z}},
\end{equation}
where $\sigma_C^y=\frac{a_0^{y,U}b_0^{y,U}a_{1}^{w,U}b_{1}^{w,U}}{a_0^{w,L}b_0^{w,L}a_1^{y,L}b_1^{y,L}}$, $\sigma_A^{z}=\frac{a_0^{z,U}a_1^{w,U}}{a_0^{w,L}a_1^{z,L}}$, $\sigma_B^{z}=\frac{b_0^{z,U}b_1^{w,U}}{b_0^{w,L}b_1^{z,L}}$, and
\begin{align}
\label{n_xx0U}\widetilde{n}_{yy}^{w, U}&=\frac{p_y}{p_w}\frac{a_0^{y,U}}{a_0^{w,L}}N_{w y}+\frac{p_y}{p_w}\frac{b_0^{y,U}}{b_0^{w,L}}N_{yw}-\frac{p_y^2}{p_w^2}\frac{a_0^{y,L}b_0^{y,L}}{a_0^{w,U}b_0^{w,U}}N_{ww},\\
\label{n_yy0l}\widetilde{n}_{zz}^{w, L}&=\frac{p_z}{p_w}\frac{a_0^{z,L}}{a_0^{w,U}}N_{w z}+\frac{p_z}{p_w}\frac{b_0^{z,L}}{b_0^{w,U}}N_{zw}-\frac{p_z^2}{p_w^2}\frac{a_0^{z,U}b_0^{z,U}}{a_0^{w,L}b_0^{w,L}}N_{ww}.
\end{align}
The detailed proof of Eq.~\eqref{N_xxL} can be found in Appendix \ref{appendix}. With Eqs.~(\ref{conclution1}, \ref{N_xxL}), we could formulate $D_{11}^L$ with only known parameters. Eq.~\eqref{N_xxL} is the most important conclusion in this paper.

With these preparation, we can now bound the fraction of counts of single-photon pulse pair among all counts caused by the signal source $z z$
\begin{equation}
  \Delta_{11}^{\prime}=\frac{n_{11}^{zz}}{N_{zz}}\ge \frac{p_z^2a_1^{z,L}b_{1}^{z,L}D_{11}^L}{N_{zz}}=\Delta_{11}^{\prime L}.
\end{equation}
Define $S_{lr}=\frac{N_{lr}}{p_l p_r N_t}$ as the yield of source $lr$. Then, we have
\begin{equation}\label{delta11}
  \Delta_{11}^{\prime L}= \frac{a_1^{z,L}b_1^{z,L}(\frac{a_1^{z,L}b_2^{z,L}\widetilde{S}_{yy}}{1+\sigma_C^y}- \frac{a_1^{y,U}b_2^{y,U}\widetilde{S}_{zz}}{1-\sigma_A^{z}-\sigma_{B}^{z}})} {a_1^{y,U}a_1^{z,L}(b_1^{y,U}b_2^{z,L}-b_2^{y,U}b_1^{z,L})S_{zz}},
\end{equation}
where
\begin{align}
\widetilde{S}_{yy}=&S_{yy}-\frac{a_0^{y,U}S_{w y}}{a_0^{w,L}}-\frac{b_{0}^{y,U}S_{yw}}{b_{0}^{w,L}}+\frac{a_0^{y,L}b_0^{y,L}S_{ww}}{a_0^{w,U}b_0^{w,U}},\\
\widetilde{S}_{zz}=&S_{zz}-\frac{a_0^{z,L}S_{w z}}{a_0^{w,U}}-\frac{b_{0}^{z,L}S_{zw}}{b_{0}^{w,U}}+\frac{a_0^{z,U}b_0^{z,U}S_{ww}}{a_0^{w,L}b_0^{w,L}}.
\end{align}

\subsection{The upper bound of the phase-flip error rate of single-photon pulse pairs}\label{errorrate}
In order to estimate the final key rate, we also need the upper bound of phase-flip error rate of single-photon pulse pair, i.e.,  $e_{11}$, which means we need to formulate the upper bound of phase-flip error counts of single-photon pulse pairs first. Here in this subsection, we only need consider the pulse pairs in $X$ basis. Similarly to Def.~2 in Sec.\ref{virtual}, we define set $H$ which contains all pulse pairs that has caused an error count and $h_{jk}$ which contains all $\ket{jk}$-photon pulse pair that has caused an error count. The probability that the $i$th pulse pair ($i\in h_{jk}$) is from each two-pulse source $lr$ $(l,r=v,x)$ is
\begin{equation}\label{probabilityError}
  Q_{i\vert jk}^{lr}=p_l p_r a_j^{l,i}b_k^{r,i}g_{jk}^i, \quad (l,r=v,x).
\end{equation}
where
\begin{equation}
  g_{jk}^{i}=\frac{1}{\sum_{l,r=v,x} {p_{l}p_{r} a_{j}^{l,i}b_{k}^{r,i}}}.
\end{equation}
If we denote the number of error counts caused by the source $lr$ as $M_{lr}$, we have
\begin{equation}\label{N_xxt}
  M_{xx}=m_{xx}^{v}+\sum_{i\in h_{11}}p_x^2a_1^{x,i}b_1^{x,i}g_{11}^i+\sum_{J}\sum_{i\in h_{jk}}p_x^2 a_j^{x,i}b_k^{x,i}g_{jk}^i,
\end{equation}
where
\begin{align}
  m_{xx}^{v}=&\sum_{k\geq 0}\sum_{i\in h_{0k}}p_x^2a_0^{x,i}b_k^{x,i}g_{0k}^i+ \sum_{j\geq 0}\sum_{i\in h_{j0}}p_x^2a_j^{x,i}b_0^{x,i}g_{j0}^i \nonumber\\
  &-\sum_{i\in h_{00}}p_x^2a_0^{x,i}b_0^{x,i}g_{00}^i.
\end{align}
If we define
\begin{equation}
  \widetilde{M}_{xx}=M_{xx}-m_{xx}^{v},
\end{equation}
Eq.~\eqref{N_xxt} can be written into
\begin{align}
  \widetilde{M}_{xx}=&\sum_{i\in h_{11}}p_x^2a_1^{x,i}b_1^{x,i}g_{11}^i+\sum_{J}\sum_{i\in h_{jk}}p_x^2 a_j^{x,i}b_k^{x,i}g_{jk}^i,\nonumber\\
  \label{N_xxtt}\ge &p_x^2a_1^{x,L}b_1^{x,L}G_{11},
\end{align}
where
\begin{equation}
  G_{11}=\sum_{i\in h_{11}}g_{11}^i.
\end{equation}
Our goal now is to formulate the upper bound of $G_{11}$. For, we have
\begin{equation}
  m_{11}^{xx}\le p_x^2a_1^{x,U}b_1^{x,U}G_{11}=m_{11}^{xx,U},
\end{equation}
where $m_{11}^{xx}$ is the number of error counts of single-photon pulse pairs for source $x x$, and $m_{11}^{xx,U}$ is the upper bound of $m_{11}^{xx}$.

With the same method to upper bound $\widetilde{N}_{yy}$, the upper bound of $G_{11}$ can be formulated by
\begin{align}
  G_{11}^{U}&=\frac{1}{a_1^{x,L}b_1^{x,L}(1-\sigma_A^{x}-\sigma_B^{x})}\left[\frac{M_{xx}}{p_x^2}- \frac{1}{p_v p_x}\frac{a_0^{x,L}}{a_0^{v,U}}M_{vx}\right.\nonumber \\
  &\left. -\frac{1}{p_v p_x}\frac{b_0^{x,L}}{b_0^{v,U}}M_{xv}+ \frac{1}{p_0^2}\frac{a_0^{x,U}b_0^{x,U}}{a_0^{v,L}b_0^{v,L}}M_{vv}\right],
\end{align}
where $\sigma_A^{x}=\frac{a_0^{x,U}a_1^{v,U}}{a_0^{v,L}a_1^{x,L}}$, $\sigma_B^{x}=\frac{b_0^{x,U}b_1^{v,U}}{b_0^{v,L}b_1^{x,L}}$. Thus, we can get the upper bound of the error rate of single-photon pulse pairs
\begin{equation}
  e_{11}=\frac{m_{11}^{xx}}{n_{11}^{xx}}\le \frac{a_1^{x,U}b_1^{x,U}G_{11}^{U}}{a_1^{x,L}b_1^{x,L}D_{11}^{L}}=e_{11}^{U}.
\end{equation}
Define $T_{\alpha\beta}=\frac{M_{lr}}{p_l p_r N_t}$ as the error yield of source $lr$. We have
\begin{equation}\label{e11}
  e_{11}^{U}=\frac{a_1^{x,U}b_1^{x,U}\widetilde{T}_{xx}}{(a_1^{x,L}b_1^{x,L})^2(1-\sigma_A^{x}-\sigma_B^{x})s_{11}^{x,L}},
\end{equation}
where
\begin{equation}
  \widetilde{T}_{xx}=T_{xx}-\frac{a_0^{x,L}T_{vx}}{a_0^{v,U}}-\frac{b_0^{x,L}T_{xv}}{b_0^{v,U}}+ \frac{a_0^{x,U}b_0^{x,U}T_{vv}}{a_0^{v,L}b_0^{v,L}},
\end{equation}
and $s_{11}^{x,L}={D_{11}^L}/{N_t}$ is the yield of single-photon pulse pair. Here we have used the fact that the lower bound of yield of  single-photon pulse pair in $X$ basis can be estimated by the lower bound of it in $Z$ basis.

\section{The final key rate and numerical simulation}\label{simulation}
In this section, we present some numerical simulations. Firstly, we shall estimate what values would be probably observed for the yields and error yields in the normal cases by the linear models~\cite{a8}. With these known values, we can calculate the lower bound of counting rate and the upper bound of phase-flip error rate of single-photon pulse pair with Eq.~\eqref{delta11} and Eq.~\eqref{e11} respectively. Then the final key rate can be calculated by
\begin{equation}
  R=S_{zz}\{\Delta_{11}^{\prime L}[1-H(e_{11}^{U})]-fH(E_{zz}) \},
\end{equation}
where $f$ is the error correction inefficiency and $H(x)=-x\log_2 x-(1-x)\log_2 (1-x)$ is the binary Shannon entropy function.

We focus on the symmetric case where the two channel transmissions from Alice to UTP and from Bob to UTP are equal. We also assume
that the UTP's detectors are identical, i.e., they have the same dark count rates and detection efficiencies, and their detection efficiencies do not depend on the incoming signals. The density matrix of the coherent state with intensity $\mu$ can be written into $\rho=\sum_{k} \frac{e^{-\mu}\mu^{k}}{k!}\oprod{k}{k}$. The actual intensity of the $i$th pulse for source $l$ out of Alice's (or Bob's) laboratory is
\begin{equation}
  \mu_{l}^{i}=\mu_{l}(1+\delta_{l}^{i}), \quad (l=v,x,w,y,z),
\end{equation}
with the boundary conditions $|\delta_{l}^{i}|\leq \delta_1$ for $l=v,w$ and $|\delta_{l}^{i}|\leq \delta_2$ for $l=x,y,z$.

\begin{figure}
\centering
\includegraphics[width=8cm]{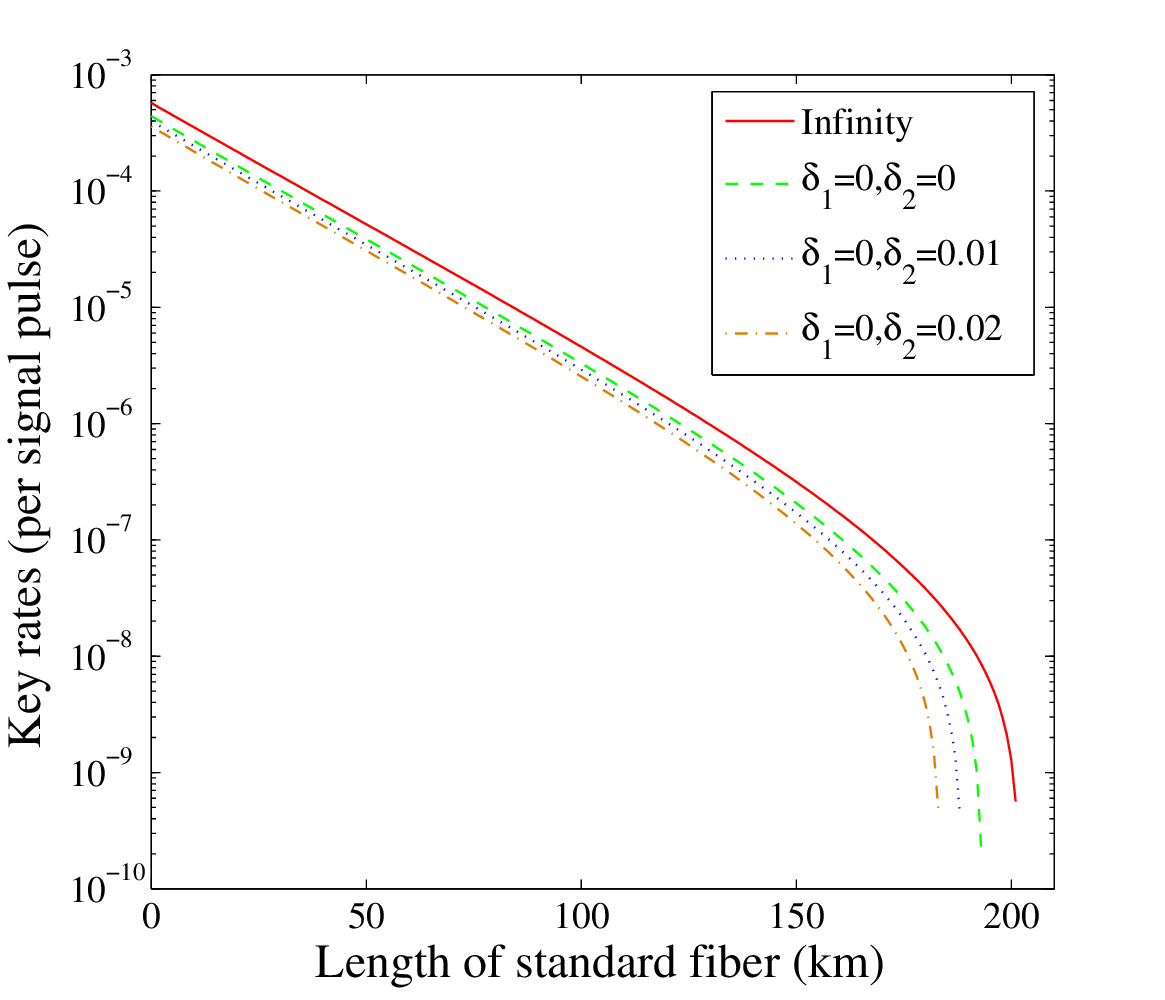}
\caption{(color online) The optimal key rates versus transmission distance (the distance between Alice and Bob). Here we set $\mu_x=0.03, \mu_y=0.03$. The intensity of the signal sources are optimized. The red solid curve is the result of the protocol with infinite number of decoy states, and other curves are the results for the protocol discussed in this work with different intensity fluctuation.}\label{figure1}
\end{figure}

\begin{figure}
\centering
\includegraphics[width=8cm]{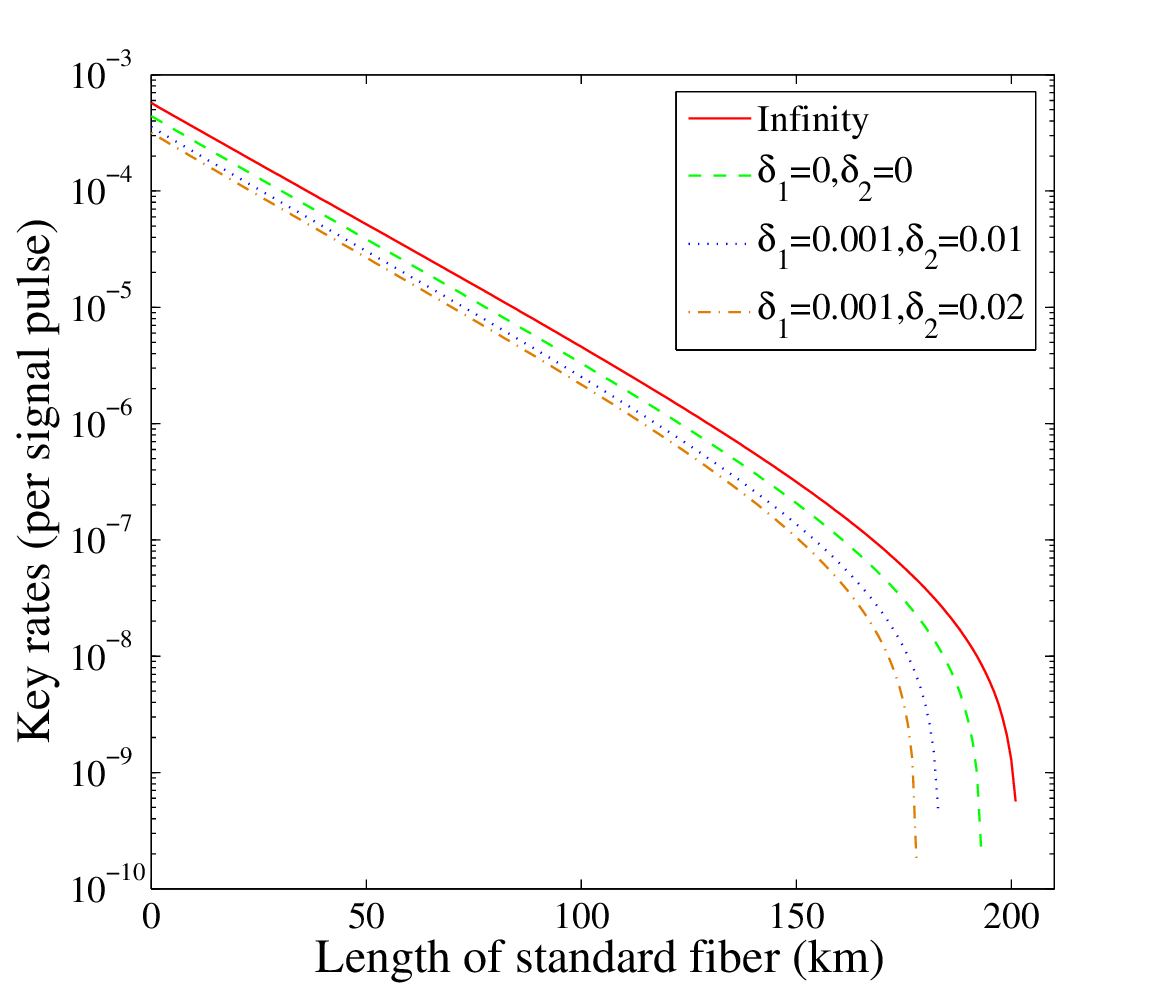}
\caption{(color online) The optimal key rates versus transmission distance (the distance between Alice and Bob). Here we set $\mu_x=0.03, \mu_y=0.03$. The intensity of the signal sources are optimized. The red solid curve is the result of the protocol with infinite number of decoy states, and other curves are the results for the protocol discussed in this work with different intensity fluctuation.}\label{figure2}
\end{figure}

Experimental conditions and the detectors properties are listed in Table~\ref{exproperty}. Fig.~\ref{figure1} and Fig.~\ref{figure2} show the key rates versus transmission distance. The red solid curve is the result of the protocol with infinite number of decoy states, and other curves are the results for the protocol discussed in this work with different intensity fluctuation.

\begin{table}
\begin{tabular}{>{\hfil}p{30pt}<{\hfil}>{\hfil}p{30pt}<{\hfil}>{\hfil}p{50pt} <{\hfil}>{\hfil}p{30pt}<{\hfil}>{\hfil}p{30pt}<{\hfil}>{\hfil}p{30pt}<{\hfil} }
\hline
\hline
$e_0$ & $e_d$ & $p_d$ & $\eta_d$ & $f$ & $\alpha_f$ \rule{0pt}{0.3cm}\\
\hline
0.5 & $1.5\%$ & $6.02\times 10^{-6}$ & $14.5\%$ & $1.16$ & $0.2$\rule{0pt}{0.3cm}\\
\hline
\hline
\end{tabular}
\caption{List of experimental parameters used in numerical simulations. $e_0$: error rate of the vacuum count. $e_d$: the misalignment-error probability. $p_d$: the dark count rate of UTP's detectors. $\eta_d$: the detection efficiency of UTP's detectors. $f$: the error correction inefficiency. $\alpha_f$: the fiber loss coefficient ($dB/km$).}\label{exproperty}
\end{table}

\section{conclusion}
In summary, we have shown how to calculate the lower bound of the fraction of single-photon pulse pair counts and the upper bound of the phase-flip error rate of the single-photon pulse pair in the decoy state MDI-QKD with source errors, provided that the parameters in the diagonal state of the source satisfy equation \eqref{decoycondition1} and \eqref{decoycondition2} and bound values of each parameters in the state is known. Our result here can be extended to the nonasymptotic case by taking statistical fluctuations into consideration in Eq.~\eqref{delta11} and Eq.~\eqref{e11}. This will be reported elsewhere.
\numberwithin{equation}{section}

{\bf{Acknowledgement:}} We thank Xiao-Long Hu and Yi-Heng Zhou for discussions. We acknowledge the financial support in part by the 10000-Plan of Shandong province (Taishan Scholars); National High-Tech Program of China Grants No. 2011AA010800 and No. 2011AA010803; National Natural Science Foundation of China Grants No. 11474182, No. 11174177, and No. 60725416; Open Research Fund Program of the State Key Laboratory of Low-Dimensional Quantum Physics Grant No. KF201513; and Key Research and Development Plan Project of ShanDong Province Grant No. 2015GGX101035.

\appendix
\section{The proof of the lower and upper bounds of $\widetilde{N}_{yy}$, $\widetilde{N}_{zz}$ and $\widetilde{M}_{xx}$} \label{appendix}
Firstly, we formulate the lower and upper bounds of $\widetilde{N}_{yy}$. The lower and upper bounds of $\widetilde{M}_{xx}$ and $\widetilde{N}_{zz}$ can be obtained similarly.

We could write $N_{wy}$,$N_{yw}$,$N_{ww}$ and $N_{yy}$ in the form of Eq.~\eqref{N_xx}
\begin{eqnarray}
  \label{N_0x} N_{wy}&=&\sum_{j,k\geq 0}\sum_{i\in c_{jk}}p_w p_y a_l^{w,i}b_m^{y,i}d_{jk}^i, \\
  \label{N_x0} N_{yw}&=&\sum_{j,k\geq 0}\sum_{i\in c_{jk}}p_w p_y a_l^{y,i}b_m^{w,i}d_{jk}^i ,\\
  \label{N_00} N_{ww}&=&\sum_{j,k\geq 0}\sum_{i\in c_{jk}}p_w^2 a_l^{w,i}b_m^{w,i}d_{jk}^{i}, \\
  \label{N_yyy}N_{yy}&=&\sum_{j,k\geq 0}\sum_{i\in c_{jk}}p_y^2 a_l^{y,i}b_m^{y,i}d_{jk}^i.
\end{eqnarray}
After a series of calculation, Eq.~\eqref{N_yyy} can be rewritten in the following equivalent form
\begin{equation}\label{N_xxvip}
  N_{yy}=\widetilde{n}_{yy}^{w}+\widetilde{N}_{yy}-A-B+C,
\end{equation}
where
\begin{eqnarray*}
  \widetilde{n}_{yy}^{w}&=&p_y^2\left[\sum_{j,k\geq 0}\sum_{i\in c_{jk}}\frac{a_0^{y,i}}{a_0^{w,i}}a_j^{w,i}b_k^{y,i}d_{jk}^i \right. \nonumber \\
  & &+\sum_{j,k\geq 0}\sum_{i\in c_{jk}}\frac{b_0^{y,i}}{b_0^{w,i}}a_j^{y,i}b_k^{w,i}d_{jk}^i \nonumber  \\
  & & \left.-\sum_{j,k\geq 0}\sum_{i\in c_{jk}}\frac{a_0^{y,i}b_0^{y,i}}{a_0^{w,i}b_0^{w,i}}a_j^{w,i}b_k^{w,i}d_{jk}^i \right],\\
  A&=&p_y^2\sum_{j,k\ge 1}\sum_{i\in c_{jk}}\frac{a_0^{y,i}}{a_0^{w,i}}a_j^{w,i}b_k^{y,i}d_{jk}^i,\\
  B&=&p_y^2\sum_{j,k\ge 1}\sum_{i\in c_{jk}}\frac{b_0^{y,i}}{b_0^{w,i}}a_j^{y,i}b_k^{w,i}d_{jk}^i,\\
  C&=&p_y^2\sum_{j,k\ge 1}\sum_{i\in c_{jk}}\frac{a_0^{y,i}b_0^{y,i}}{a_0^{w,i}b_0^{w,i}}a_j^{w,i}b_k^{w,i}d_{jk}^i.
\end{eqnarray*}
One may easily prove that
\begin{equation*}
  n_{yy}^{w}=\widetilde{n}_{yy}^{w}-A-B+C,
\end{equation*}
where
\begin{eqnarray*}
  n_{yy}^{w}&=&\sum_{k\geq 0}\sum_{i\in c_{0,k}}p_y^2a_0^{y,i}b_k^{y,i}d_{0k}^i+\sum_{j\geq 0}\sum_{i\in c_{j,0}}p_y^2a_j^{y,i}b_0^{y,i}d_{j0}^i\nonumber\\
  & &-\sum_{i\in c_{0,0}}p_y^2a_0^{y,i}b_0^{y,i}d_{00}^i.
\end{eqnarray*}
Given Eqs.~(\ref{N_0x}-\ref{N_00}), the lower bound and upper bound of $\widetilde{n}_{yy}^{w}$ could be formulated as
\begin{align*}
  \widetilde{n}_{yy}^{w,L}&=\frac{p_y}{p_w}\frac{a_0^{y,L}}{a_0^{w,U}}N_{wy}+\frac{p_y}{p_w}\frac{b_0^{y,L}}{b_0^{w,U}}N_{yw}- \frac{p_y^2}{p_w^2}\frac{a_0^{y,U}b_0^{y,U}}{a_0^{w,L}b_0^{w,L}}N_{ww},\\
  \label{n_xx0uu}
  \widetilde{n}_{yy}^{w,U}&=\frac{p_y}{p_w}\frac{a_0^{y,U}}{a_0^{w,L}}N_{wy}+\frac{p_y}{p_w}\frac{b_0^{y,U}}{b_0^{w,L}}N_{yw}- \frac{p_y^2}{p_w^2}\frac{a_0^{y,L}b_0^{y,L}}{a_0^{w,U}b_0^{w,U}}N_{ww}.
\end{align*}
It is easy to know that the lower bound of $A,B$ and $C$ is just
\begin{equation}
  A^L=0,\quad B^L=0,\quad C^L=0.
\end{equation}
The upper bound of $A,B$ and $C$ could be formulated as follows
\begin{eqnarray*}\label{C}
  C&=&p_y^2\sum_{j,k\geq 1}\sum_{i\in c_{jk}}\frac{a_0^{y,i}b_0^{y,i}}{a_0^{w,i}b_0^{w,i}}a_j^{w,i}b_k^{w,i}d_{jk}^i,  \\
  &=&p_y^2\sum_{j,k\geq 1}\sum_{i\in c_{jk}}\sigma_1^U a_j^{w,i}b_k^{w,i}d_{jk}^i-\zeta_1 \nonumber \\
  &=&p_y^2\sum_{j,k\geq 1}\sum_{i\in c_{jk}}\sigma_1^U \frac{a_1^{w,i}b_1^{w,i}}{a_1^{y,i}b_1^{y,i}}a_j^{y,i}b_k^{y,i}d_{jk}^i-\zeta_1-\zeta_2 ,\nonumber \\
  &=&p_y^2\sum_{j,k\geq 1}\sum_{i\in c_{jk}}\sigma_1^U \sigma_2^U a_j^{y,i}b_k^{y,i}d_{jk}^i-\zeta_1-\zeta_2-\zeta_3, \nonumber
\end{eqnarray*}
where
\begin{align*}
  \sigma_1^U&=\frac{a_0^{y,U}b_0^{y,U}}{a_0^{w,L}b_0^{w,L}},\quad \sigma_2^U=\frac{a_1^{w,U}b_1^{w,U}}{a_1^{y,L}b_1^{y,L}},\\
  \zeta_1&=p_y^2\sum_{j,k\ge 1}\sum_{i\in c_{jk}}(\sigma_1^U-\frac{a_0^{y,i}b_0^{y,i}}{a_0^{w,i}b_0^{w,i}})a_j^{w,i}b_k^{w,i}d_{jk}^i\ge 0,\\
  \zeta_2&=p_y^2\sum_{j,k\ge 1}\sum_{i\in c_{jk}}\sigma_1^U(\frac{a_1^{w,i}b_1^{w,i}}{a_1^{y,i}b_1^{y,i}}a_j^{y,i}b_k^{y,i}-a_j^{w,i}b_k^{w,i})d_{jk}^i, \\
  \zeta_3&=p_y^2\sum_{j,k\ge 1}\sum_{i\in c_{jk}}\sigma_1^U(\sigma_2^U-\frac{a_1^{w,i}b_1^{w,i}}{a_1^{y,i}b_1^{y,i}})a_j^{y,i}b_k^{y,i}d_{jk}^i\ge 0,
\end{align*}
$\zeta_2\ge 0$ can be directly obtained with Eq.~\eqref{decoycondition2}. We have
\begin{equation}\label{C_U}
  C^U=\sigma_1^U\sigma_2^U\widetilde{N}_{yy}.
\end{equation}
Similarly, the upper bounds of $A$ and $B$ are
\begin{equation}\label{A_U}
  A^U=\sigma_A^{y}\widetilde{N}_{yy},\quad B^U=\sigma_B^{y}\widetilde{N}_{yy},
\end{equation}
where
\begin{equation}\label{sigmaAB}
  \sigma_A^{y}=\frac{a_0^{y,U}a_1^{w,U}}{a_0^{w,L}a_1^{y,L}} ,\quad  \sigma_B^{y}=\frac{b_0^{y,U}b_1^{w,U}}{b_0^{w,L}b_1^{y,L}} .
\end{equation}
With Eq.~\eqref{N_xxvip}, we have
\begin{equation}\label{N_xxl1}
  \widetilde{N}_{yy}=N_{yy}-\widetilde{n}_{yy}^w+A+B-C\ge N_{yy}-\widetilde{n}_{yy}^{w,U}-C^U,
\end{equation}
and
\begin{equation}\label{N_xxu1}
  \widetilde{N}_{yy}\le N_{yy}-\widetilde{n}_{yy}^{w,L}+A^U+B^U.
\end{equation}
Combining Eq.~\eqref{C_U} with Eq.~\eqref{N_xxl1}, we can formulate $\widetilde{N}_{yy}^L$ as follows
\begin{equation}\label{N_xxll}
  \widetilde{N}_{yy}^L=\frac{N_{yy}-\widetilde{n}_{yy}^{w,U}}{1+\sigma_C^y},
\end{equation}
where $\sigma_C^y=\sigma_1^U\sigma_2^{U}$. \\
Combining Eq.~\eqref{A_U} with Eq.~\eqref{N_xxu1}, we can formulate $\widetilde{N}_{yy}^U$ as
\begin{equation}\label{N_xxu}
  \widetilde{N}_{yy}^U=\frac{N_{yy}-\widetilde{n}_{yy}^{w,L}}{1-\sigma_A^{y}-\sigma_B^{y}},\\
\end{equation}
where $\sigma_{A}^{y}$ and $\sigma_{B}^{y}$ have been defined in Eq.~\eqref{sigmaAB}.

By using the same way, we can formulate the lower and upper bounds of $\widetilde{N}_{zz}$ and $\widetilde{M}_{xx}$. Actually, we only need the upper bounds of $\widetilde{N}_{zz}$ and $\widetilde{M}_{xx}$. Explicitly, we have
\begin{equation}
  \widetilde{N}_{zz}=\frac{N_{zz}-\widetilde{n}_{zz}^{w,L}}{1-\sigma_{A}^z-\sigma_B^z}, \quad \widetilde{M}_{xx}=\frac{M_{xx}-\widetilde{m}_{xx}^{v,L}}{1-\sigma_{A}^x-\sigma_B^x},
\end{equation}
where
\begin{align*}
  \widetilde{n}_{zz}^{w,L}&=\frac{p_z}{p_w}\frac{a_0^{z,L}}{a_0^{w,U}}N_{wz}+\frac{p_z}{p_w}\frac{b_0^{z,L}}{b_0^{w,U}}N_{zw}-\frac{p_z^2}{p_w^2}\frac{a_0^{z,U}b_0^{z,U}}{a_0^{w,L}b_0^{w,L}}N_{ww},\\
  \widetilde{m}_{xx}^{v,L}&=\frac{p_x}{p_v}\frac{a_0^{x,L}}{a_0^{v,U}}M_{vx}+\frac{p_x}{p_v}\frac{b_0^{x,L}}{b_0^{v,U}}M_{xv}-\frac{p_x^2}{p_v^2}\frac{a_0^{x,U}b_0^{x,U}}{a_0^{v,L}b_0^{v,L}}M_{vv},
  \end{align*}
  and
\begin{align*}
  \sigma_A^{z}&=\frac{a_0^{z,U}a_1^{w,U}}{a_0^{w,L}a_1^{z,L}},\quad \sigma_B^{z}=\frac{b_0^{z,U}b_1^{w,U}}{b_0^{w,L}b_1^{z,L}}, \\
  \sigma_A^{x}&=\frac{a_0^{x,U}a_1^{v,U}}{a_0^{v,L}a_1^{x,L}},\quad \sigma_B^{x}=\frac{b_0^{x,U}b_1^{v,U}}{b_0^{v,L}b_1^{x,L}}.
\end{align*}

\end{document}